\documentclass[aps,prl,twocolumn,floats,graphicx,epsfig,euscript,amsmath,amssymb]{revtex4}
\usepackage[dvips]{graphicx}
\usepackage{epsfig}
\DeclareMathAlphabet\EuScript{U}{eus}{m}{n} \SetMathAlphabet\EuScript{bold}{U}{eus}{b}{n}

\def\lapprox{\,\raise0.4ex\hbox{$<$}\kern-0.8em\lower0.7ex\hbox{$\sim$}\,}
\def\gapprox{\,\raise0.4ex\hbox{$>$}\kern-0.8em\lower0.7ex\hbox{$\sim$}\,}
\begin{document}
\title
{Super-long life time for 2D cyclotron spin-flip excitons}

\author{$\qquad$ L.V. Kulik, A.V. Gorbunov, A.S. Zhuravlev,~V.B.~Timofeev,
$\qquad\qquad\qquad{}\qquad\qquad$ S. Dickmann,
and I.V. Kukushkin$\quad \qquad\qquad$}

\affiliation{$$Institute of Solid State Physics, Russian
Academy of Sciences, Chernogolovka, 142432 Russia}

\date{\today}

\begin{abstract}
An experimental technique for the indirect manipulation and detection of electron spins entangled in two-dimensional magnetoexcitons has been developed. The kinetics of the spin relaxation has been
investigated. Photoexcited spin-magnetoexcitons were found to exhibit extremely slow relaxation
in specific quantum Hall systems, fabricated in high mobility GaAs/AlGaAs structures; namely, the
relaxation time reaches values over one hundred microseconds. A qualitative explanation of this
spin-relaxation kinetics is presented. Its temperature and magnetic field dependencies are discussed
within the available theoretical framework.

\vskip 1mm

\noindent PACS numbers: 73.43.Lp,71.70.Di,75.30.Ds
\end{abstract}
\maketitle
The problem of spin relaxation in solids remains the main focus of research related to spintronics
and the search for long-lived non-equilibrium spin systems that could be promising for the realization of quantum calculations. A long lifetime of the excited quantum states in two-level systems is
an essential condition for the practical realization of quantum information bits (qubits). The majority of research aimed at creating solid-state qubits until now has been focused on issues of electron spin relaxation in quantum-confined structures: quantum dots \cite{Hanson07}, impurity centers \cite{Gaebel}, and strongly correlated
quasi-one-dimensional systems related to chiral Hall currents \cite{Karzig}. We present the results of our research
into super-long spin relaxation times for a translation-invariant (microscopically delocalized) quantum
object, the cyclotron spin-flip exciton (CSFE), which can be manipulated using the photoexcitation of two-dimensional (2D) electrons.

It is known that exciton states in metal systems, such as a 2D electron gas, are unstable \cite{Cui}. However,
2D metal becomes an insulator for integer quantum Hall states in quantizing magnetic fields at temperatures
significantly lower than the cyclotron energy. Neutral excitations in quantum Hall insulators are
magnetoexcitons (bosons) by analogy with the magnetoexcitons in 3D semiconductor systems \cite{Gor'kov}. Indeed,
the analogy applies if a cyclotron, Zeeman or Coulomb gap replaces the forbidden energy gap in semiconductors.
The occurrence of a bosonic component in a correlated fermion system makes the quantum
Hall states a potential candidate for the formation of macroscopic, nonequilibrium condensates, demonstrating superfluidity as suggested in \cite{lerner}.

One of the simplest realizations is the magnetoexciton induced by an electron promoted from the
zeroth to the first Landau level in a spin-unpolarized quantum Hall system at a filling factor of $\nu\!=\!2$. The
excitation spectrum exhibits two types of inter-Landau-level magnetoexcitons: a spin-singlet with a total
spin of $S\!=\!0$ possessing an energy equal to the cyclotron gap (according to the Kohn theorem \cite{Kohn,Kallin}); and a
CSFE spin-triplet with a total spin of $S\!=\!1$ and spin projections of $-1$, 0, and +1 along the magnetic
field axis \cite{Kallin}. The singlet magnetoexciton represents the spinless magnetoplasma mode, and its relaxation
is related to the transition of an electron from the first to the zeroth Landau level. This process may occur via dipole cyclotron radiation, and this radiative mechanism is the dominant relaxation channel in this case \cite{Zhang}. In other words, the spin-singlet exciton is a `bright exciton'. In contrast to the spin-singlet,
the spin-triplet exciton is not radiatively active due to electron spin conservation. Thus, the spin-triplet
exciton is a dark exciton. Additionally, there are no symmetry restrictions similar to the Kohn theorem
for the spin-triplet exciton, and its energy is reduced relative to the cyclotron energy due to the Coulomb interaction \cite{Kulik,di05}. An excitation of the lowest-energy 
CSFE component $S\!=\!1$, $S_z\!=-\!1$ will change the spin state of the
electron system (see diagram in Fig. 1), therefore the CSFE relaxation also is the spin relaxation for the entire electron system.

The optical excitation of electrons to higher energy states can be used to produce a significant
spin-triplet exciton density. The relaxation of the spin-triplet exciton to the ground state accompanied by electron spin flip is expected to be a long-term process \cite{Dickmann}. 
If the electron system temperature is much
lower than the energy gap between the spin-singlet and spin-triplet excitons, the latter 
has no optical
decay channels, and the only nonradiative channel for spin-triplet exciton relaxation to 
ground state is
the emission of high-frequency acoustic phonons in the presence of spin-orbit coupling. 
The present
paper aims at generating the $S\!=\!S_z\!=\!1$ exciton ensemble and investigates the 
relaxation of this ensemble to the ground state.

A high electron mobility of the 2D system is essential for
spin-triplet exciton observation. Therefore, we considered
high-quality heterostructures with symmetrically doped GaAs/AlGaAs
single quantum wells. Optical measurements were made using two 
continuous wave tunable lasers (Fig. 1, top panel): one for resonant 
excitation of the electron system and the other for recording the spectra of resonance
reflectance (RR), photoluminescence (PL), and inelastic light scattering 
(ILS). Because detailed spectral reflectance profile analysis in doped systems can be fairly complicated \cite{Shields}, it can be postulated that for
the $\nu\!=\!2$ quantum Hall insulator, the reflectance signal of a laser tuned 
to the transition from the zeroth Landau level of heavy valence holes to the zeroth Landau level of conductance band electrons will not be observed, because all of the electron states are occupied. With a supplementary pump that promotes the electrons to higher Landau levels, one can expect the formation of dark excitons, $S\!=\!1,\;S_z\!=\!-1$,   as
they occupy the lowest energy excited state in which the pumped carriers 
relax (considering the negative value of the electron g-factor in GaAs, 
$S\!=\!S_z\!=\!1$). This process should manifest itself as a decrease in
the unoccupied states on the first electron Landau level and the occurrence 
of empty states (vacancies) on the zeroth electron Landau level. The 
respective change in RR is called the photoinduced resonance reflection 
(PRR).
\vspace{-0.mm}
\hspace{-0cm}
\begin{figure}[htb!]
\hspace{-25.7mm}
\includegraphics[scale=1.1,clip]{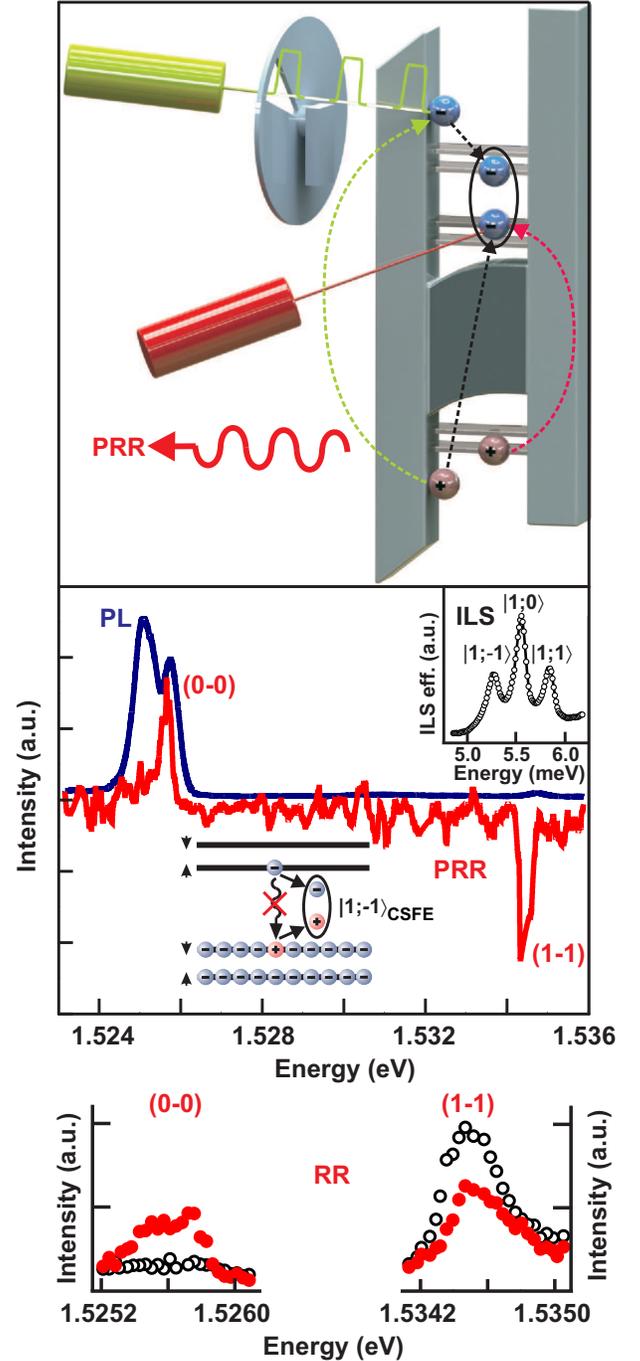}
\vspace{-6mm} \caption{\label{fig1} Photoluminescence (PL) and photoinduced resonant reflectance (PRR) spectra taken at  $\nu\!=\!2$  in
a $17\,$nm quantum well (the dark mobility is $5\!\times\!10{}^6$cm${}^2$/Vs, the electron concentration is $2.4\!\times\!10^{11}$cm${}^{-2}$)
for the magnetic field  $H\!=\!5\,$T 
normal to the quantum well and bath temperature of $0.45\,$K. The diagram
shows how the lowest energy excited state forms. 
The inset demonstrates the inelastic light scattering (ILS)
spectrum of the cyclotron spin-flip exciton taken at similar 
experimental conditions (perpendicular magnetic
field and temperature) with a $5\,$T extra-parallel magnetic 
field to enhance the Zeeman energy splitting above
the experimental spectral resolution. Detailed resonant 
reflectance (RR) spectra with (closed dots) and
without (open dots) extra pumping are shown in the bottom. 
(top) Scheme of the experimental setup.
\vspace{-8mm}}
\end{figure}

The PRR spectrum should have two peaks: a positive peak in the transition region from the 
zeroth
Landau level of heavy holes to the upper spin sublevel of the zeroth electron Landau level 
and a negative
peak in the transition region from the first valence heavy-hole Landau level to the first 
electron Landau
level. The positive peak is responsible for the formation of vacancies on the upper spin 
sublevel of the
zeroth electron Landau level, while the negative peak is responsible for the decrease in 
the number of
vacancies on the lower spin sublevel of the first electron Landau level. Thus, we propose 
a photoinduced
resonance reflectance technique for indirectly testing the presence of dark excitons 
through optically
allowed transitions from the valence band to the conductance band.

{\em Results.} The central panel of Fig.  1 shows the representative experimental PL and PRR spectra taken at $\nu\!=\!2$.
The detailed RR spectra with and without extra pumping are presented in the bottom of Fig.  1. The
PRR spectrum looks just as it was anticipated: a positive peak at the (0-0) 
transition line (the first figure
refers to the valence heavy-hole Landau level and the second to the 
electron Landau level) and a negative
peak in the region of (1-1) transition. This pattern is observed in the 
experiment, which is related to the
formation of low energy magnetoexcitons composed of electrons on the 
first Landau level and vacan -cies on the zeroth electron Landau level 
(see the diagram in Fig.  1). Despite the inactivity of cyclotron
spin-triplet excitons with respect to radiative decay, their actual 
existence is established by ILS, and the
shift of the central triplet component  $S\!=\!1,\,\;S_z\!=\!0$,
from the cyclotron energy is used to determine the
energy gap between the spin-triplet and spin-singlet excitons 
(Fig. 1, inset)

\begin{figure}[htb!]
\vspace{-3mm}
\includegraphics[scale=1,clip]{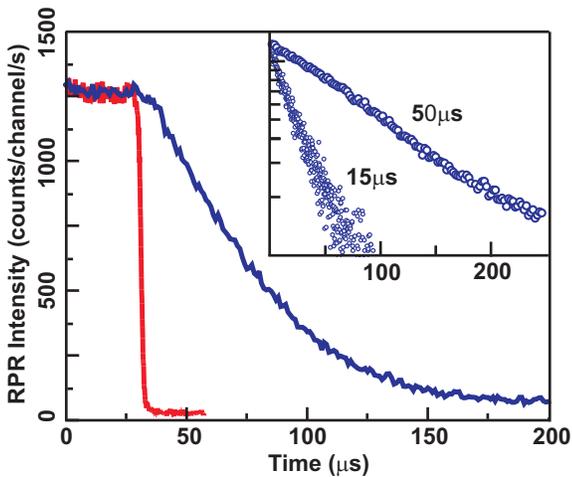}
\vspace{-4mm}\caption{\label{fig2} Example of a
photostimulated resonant reflectance (PRR) decay curve with the
apparatus function of the chopper (red). The inset shows two decay
kinetics for quantum wells of 35~nm (longer) and 17~nm (shorter)
width taken at $\nu=2$ at 4~T perpendicular magnetic field
and bath temperature of 0.4~K.\vspace{-3.5mm}}
\end{figure}

The pumping laser emission was modulated periodically to measure the 
magnetoexciton relaxation kinetics (Fig. 2). The PRR decay curve on the 
(0-0) transition line appears to be extremely long. The PRR
signal decays over tens of microseconds upon switching off the resonant 
pumping (Fig. 2). The PRR (1-1) signal in turn increases over the same 
time, which points to common relaxation dynamics for the exciton
states formed by electrons on the first and by holes on the zeroth Landau 
level. The common specific feature of the observed relaxation kinetics is 
their temperature dependence. The temperature dependence for the spin 
relaxation rate is exponential at high temperatures above $1\,$K, with a 
characteristic relaxation
time ($\tau_1$) on the order of 1 nanosecond and an energy gap ($\Delta$) of 
approximately $11\,$K (Fig. 3). This is
indicative of an optically activated relaxation channel consisting of an 
electron spin-flip due to spin-orbit
coupling, increasing the exciton energy to the cyclotron energy, and 
emitting a photon with the cyclotron energy \cite{Zhang}.
We have measured the Coulomb term independently for the sample in question, and it
amounts to $0.54\,$meV; hence, the Coulomb gap is $6.3\,$K. We should add the Zeeman term for electrons
to this quantity (an electron should rotate its spin when relaxing to 
the ground state). This adds $1.2\,$K
at $4\,$T to the Coulomb term. We have a total of $7.5\,$K, which is somewhat less than the obtained $11\,$K.

The kinetics are no longer temperature dependent at lower temperatures, 
i.e., the relaxation mechanism changes. Temperature-independent relaxation 
implies a non-activated mechanism for the exciton decay. The most 
intuitive relaxation channel could be represented by the decay into 
short-wave acoustic phonons described in Ref. \cite{Dickmann}.
The relaxation time  $t_0$ is predicted to depend superlinearly on the extent of the
electron wave function in the direction of quantum well growth (in the  
z-direction). To estimate this effect, two types of quantum wells, 17 and 
35 nm wide, with nearly
equal electron densities were chosen so that the half-widths of their 
electron envelope wave functions had approximately a two-fold difference.
Fig. 3 confirms qualitatively the main prediction of Ref. \cite{Dickmann}.
This Figure supports also another prediction of Ref. \cite{Dickmann}
the relaxation rate falls with magnetic field, as the emission of hard phonons with large frequencies ($\sim\omega_c$) is difficult due to reduction of the electron-phonon 
coupling.

{\em Discussion}. We will now discuss the features of the
experimental data obtained. (i) The CSFE relaxation is extremely slow,
which, in principle, corresponds to the general theoretical
prediction. However, the characteristic times (Fig. 3) still
differ significantly from the 7~ms calculated for CSFE
annihilation at a high CSFE density \cite{Dickmann} (see the
time $\tau$ therein). (ii) With a growing magnetic field, i.e., with
an increasing cyclotron gap, the relaxation becomes appreciably
slower (Fig. 3), which excludes the radiation mechanism (the latter
should, on the contrary, be enhanced owing to the dipole irradiation law 
$\sim\omega_c^3$) and suggests a phonon-emission
relaxation process. (iii) Relaxation in a wider quantum well
occurs much more slowly (see the illustration in Fig.~3) which is also
a qualitative indication of a phonon-emission relaxation channel
(the larger the extent  of the electron wave function in the
$z$-direction, the smaller the effective electron coupling to a
short-wave acoustic phonon). (iv) Relaxation does not depend on the 
temperature below a
characteristic temperature,  which 
is consistent with an acoustic phonon relaxation mechanism. (v)
Finally, within our experimental accuracy, relaxation occurs
exponentially with time, i.e., the relaxation rate is independent
of the CSFE concentration.

The  last experimental observation contradicts
the CSFE--CSFE scattering relaxation scenario \cite{Dickmann}, which
assumes a single-phonon emission resulting in the annihilation of one
of the two excitons, and therefore leading to a non-exponential
time dependence of the relaxation rate. Additionly, the fact that the
observed relaxation times are still much shorter than those
predicted forces us to correct the existing theoretical
assumptions. Recall that the extremely long CSFE relaxation is
essentially due to the small factor
$|\Phi(k_z)|^2=|\int\!e^{ik_zz}|\chi(z)|^2 dz|^2$, where $\chi(z)$
is the 2D electron size-quantized wave function  and $k_z\approx
k_{\rm phonon}\approx \omega_c/c_l$ ($\omega_c$ to denote the cyclotron frequency, $c_l$ is the longitudinal sound velocity). Yet, the energy release via phonon emission could
be a more complicated process than previously thought, i.e.,
two- or even multi-phonon emission channels may occur.
\begin{figure}[htb!]\vspace{-0mm}
\includegraphics[scale=1,clip]{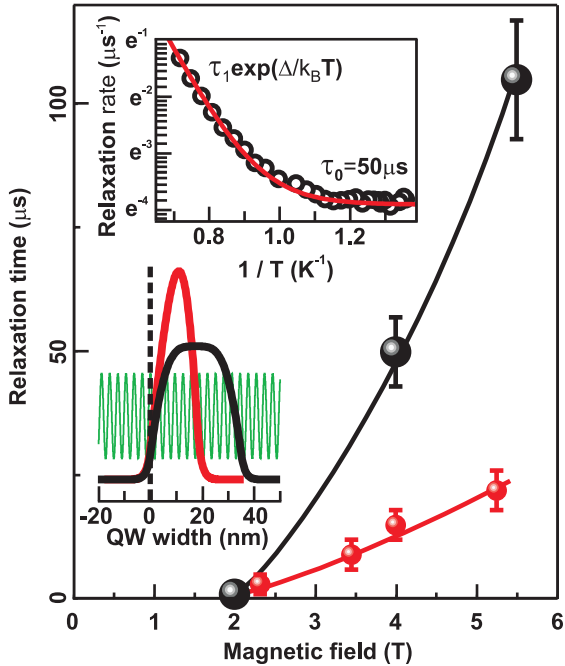}
\vspace{-3mm} \caption{\label{fig3} An assembled graph of low temperature relaxation 
constant $\tau_0$ as a function of magnetic field.
The small points denote the magnetic field dependence for quantum wells of $17\,$nm widths. The large points
denote the magnetic field dependence for quantum wells of $35\,$nm widths. The curves are 
drawn for clarity.
Top inset: relaxation rate vs. temperature in logarithmic (ln) scale taken at $\nu\!=\!2$ 
in the $35\,$nm quantum
well at $4\,$T magnetic field (dots) with a fitting using two relaxation mechanisms, 
activated and temperature-independent (line).The diagram shows the electron size quantized envelopes for
two quantum wells of 17 and $35\,$nm widths and a schematic wave function of the phonon 
emitted
to illustrate an incommensurability
between the phonon wave length and the extent of the electron wave function in the growth 
direction of the quantum wells.
\vspace{-5mm}}
\end{figure}

Let us consider for example the {\em two-phonon} emission
mechanism. It may be realized as a {\em single-exciton}
annihilation process representing a transition from the initial CSFE
state to the final state with two longitudinal 3D phonons with
wave vectors ${\bf k}_1\!=\!(-{\bf q},k_{z1})$ and  ${\bf
k}_2\!=\!({\bf q},k_{z2})$. In-plane wave-vector components
with $q\sim 1/l_B$ ($l_B$ is the magnetic length) provide the most effective
electron--phonon coupling in a strong perpendicular magnetic field, whereas 
due to energy conservation, the
$z$-components must satisfy the
condition $|k_{1z}|\!+\!|k_{2z}|\!\approx\!\omega_c/c_l\gg 1/d$
($d$ is the characteristic extent of the electron wave function in
the $z$-direction).
For two-phonon emission processes, the small vertex of the
electron--phonon interaction is perturbatively accounted for
within the second order approximation (whereas the spin--orbit
coupling is still considered to the first order). However, the
two-phonon relaxation channel could be faster than the
single-phonon one because (i) the phonon phase volume for emitted
phonons becomes larger: now the momentum conservation law allows
phonons to have anti-parallel components $q\!\sim\!1/l_B$ ($l_B$ is the magnetic length)
effectively contributing to the relaxation; (ii) the relevant two-phonon 
transition matrix element is enhanced because it is determined by 
numerous intermediate virtual-exciton states corresponding to an electron
promoted to various Landau levels with numbers $n\!\geq\!1$; (iii) finally, the
product $|\Phi(k_{1z})|^2|\Phi(k_{2z})|^2$ enters the final result for the 
relaxation rate rather than the factor $|\Phi(\omega_c/c_l)|^2$. Appropriate 
analysis reveals that
the most effective relaxation is realized via the emission of phonons
with $|k_{1z}|\!\approx\!|k_{2z}|\approx\!\omega_c/2c_l$, and the factor $|\Phi(k_{1z})|^2|\Phi(k_{2z})|^2$, may be much larger than $|\Phi(\omega_c/c_l)|^2$
as a result. Similar arguments are
valid for a multi-phonon process contributing to the
single-exciton annihilation. Therefore, it is no wonder that the averaged
single-exciton relaxation matrix element may exceed the
double-exciton one theoretically studied in Ref. \cite{Dickmann}.

In conclusion, we have created an ensemble of 2D translation-invariant cyclotron spin-flip 
excitons and investigated its relaxation kinetics. The spin-excitons (spin waves)
associated with a spin flip within
a single Landau level have been known thus far to possess the longest relaxation times of approximately 100 ns measured among translation-invariant spin excitations 
in 2D-electron systems \cite{Zhuravlev}. The studied
spin-triplet exciton associated with electron spin flip and the simultaneous change of 
cyclotron quantum
number now relaxes to the ground state over tens or even hundreds of microseconds. 
The unprecedentedly large relaxation times for excitations in the translation-invariant system are partly due to a specific Coulomb confinement in the conjugate $K$-space that blocks some relaxation channels and thereby
prevents excited electrons from backward spin flipping. Coulomb confinement is somewhat 
similar to
spatial 3D confinement forming quantum dots and also resulting in the slowdown of 
single-electron spin relaxation \cite{Hanson07}.
Owing to such long relaxation times, optical pumping can be used to generate cyclotron
spin-flip excitons with densities sufficient to reveal their collective Bose properties. In 
case of successful realization of this scenario the long living cyclotron spin-flip exciton ensemble may appear as a second example, after electron-electron bilayers \cite{Eisenstein}, 
of a dense system of Bose particles in degenerate 2D Fermi
gas, which may demonstrate Bose-Einstein condensation effects.

{\em Methods}. We explored two sets of high-quality heterostructures (the dark mobility in 
the range of $5–20\!\times\!10^6$cm${}^2/$Vs) with symmetrically doped GaAs/AlGaAs single 
quantum wells of two widths: 17 and
35  nm. Every set included a number of samples with different electron concentrations in the 2D channel
ranging from $5\!\times\!10^{10}$ to $2.5\!\times\!10^{11}\,$cm${}^{-2}$.
Measurement of every sample characterized by its unique elec -tron concentration gave us one point in the graph of spin relaxation time versus magnetic field.

The sample of size of approximately $3\!\times\!3\,$mm${}^2$
was placed into a pumped cryostat with liquid
${}^3$He, which in turn was placed into a
${}^4$He cryostat with a superconducting solenoid.
The setup allowed measurements at the bath temperature down to $0.45\,$K and at the 
magnetic field up to $14\,$T. Optical measurements were made using the dual-fiber 
technique with the use of multimode quartz
glass optical fiber with a core diameter of $400\,\mu$m and numerical aperture N. A. of 
0.39. One fiber was
used for photoexcitation, and a second for collecting the emission from the sample and 
transferring it
onto the entrance slit of a grating spectrometer equipped with a CCD camera cooled by 
liquid nitrogen.
Two continuous wave tunable lasers with narrow spectral widths of emission lines 
(20 and $5\,$MHz) were
employed as optical sources, which enabled us to use one of the lasers for resonant 
excitation of the
electron system and the other for recording spectra of resonance reflectance, 
photoluminescence, and
inelastic light scattering.

The experimental geometry was chosen when measuring the resonance reflectance spectra so that the
specularly reflected beam axis coincided with the receiving fiber axis at an angle of incidence of approximately $10^{\rm o}$. The contribution of the sample surface reflection was suppressed using crossed linear polarizers
set between the sample and the ends of the pumping and collecting fibers. One of the lasers was used as an
optical pump of the system via the excitation of the electrons to high energy Landau levels,
$n\!>\!1$, in the
experimental studies of photo-induced reflection. The pumping laser intensity was limited to a power below $0.3\,$mW to minimize heating effects. An emission from a probing laser, which was weaker by an order of
magnitude, was coupled into the same waveguide. The resonance reflectance spectrum was obtained by
scanning the emission wavelength of the probing laser and registering the laser line intensity using the
spectrometer with the CCD camera. The photoinduced resonant reflection was obtained as the difference
in the resonance reflectance spectra with the resonance pump switched on and off.

The pumping laser emission was modulated periodically using a mechanical
chopper (a rotating disk with a radial slit), in the experiment on photoinduced reflection 
kinetics. The modulation period was approximately $11\,$ms, and the pump pulse rise/decay 
time was less than $2\,\mu$s. The laser beam was focused onto the chopper disk surface by 
a microscope objective to shorten the pump pulse front. The excitation energy of the 
probing laser was set to the maximum/minimum of the
photoinduced resonance reflectance spectrum to record the decay/rise curve, respectively. 
The probing laser emission reflected from the sample was passed through a tunable 
narrow-band interference filter of
1.1 nm spectral width to cut off the pumping laser light and then focused on an avalanche photodiode
operating in the photon counting regime. A gated photon counter was used to accumulate the 
photoinduced resonant reflection signal as a function of the time delay from the pump 
cut-off event to obtain a
photoinduced resonant reflection decay curve with an acceptable signal-to-noise ratio.

The authors acknowledge the Russian Fund of Basic Research for
support.

\end{document}